\documentclass[%
 reprint,
 amsmath,amssymb,
 aps,
]{revtex4-2}

\usepackage{graphicx}
\usepackage{dcolumn}
\usepackage{bm}

\usepackage{subcaption}
\usepackage{tikz}
\usetikzlibrary{shapes.geometric,arrows,positioning,fit,backgrounds}
\tikzstyle{startstop} = [rectangle, rounded corners, 
minimum width=1.5cm,
minimum height=0.75cm,
text centered,
font = {\small},
draw=black]

\tikzstyle{conv} = [rectangle, 
trapezium stretches=true, 
minimum width=1.5cm,
minimum height=0.75cm,
text centered, 
font = {\small},
draw=black, 
fill=blue!30]

\tikzstyle{sepconv} = [rectangle, 
trapezium stretches=true, 
minimum width=1.5cm,
minimum height=0.75cm,
text centered, 
font = {\small},
draw=black, 
fill=blue!10]

\tikzstyle{dense} = [rectangle, 
minimum width=1.5cm,
minimum height=0.75cm,
text centered, 
font = {\small},
draw=black, 
fill=orange!30]

\tikzstyle{concat} = [rectangle, 
minimum width=1.5cm,
minimum height=0.75cm,
text centered, 
font = {\small},
draw=black, 
fill=gray!20]

\tikzstyle{pooling} = [rectangle, 
minimum width=1.5cm,
minimum height=0.75cm,
text centered, 
font = {\small},
draw=black, 
fill=red!30]

\tikzstyle{arrow} = [thick,->,>=stealth]

\tikzset{
  arlabel/.style={font=\scriptsize, align=left, inner sep=1pt, fill=white},
  subconv/.style={conv, font=\scriptsize, minimum height=0.55cm}
}

\tikzset{
  arlabel/.style={font=\scriptsize, align=left, inner sep=1pt, fill=white}
}

\begin{document}

\preprint{APS/123-QED}

\title{On-chip probabilistic inference for charged-particle tracking at the sensor edge}
\thanks{FERMILAB-PUB-26-0100-CSAID-ETD}%

\author{Jennet Dickinson}
\author{Benjamin Weiss}
\affiliation{Cornell University, Ithaca, NY 14853, USA}

\author{Doug Berry}
\author{Giuseppe Di Guglielmo}
\altaffiliation[Also at ]{Northwestern University, Evanston, IL 60208, USA}
\author{Farah Fahim}
\altaffiliation[Also at ]{Northwestern University, Evanston, IL 60208, USA}
\author{Abhijith Gandrakota}
\author{Lindsey Gray}
\email{Corresponding author. Email: \url{lagray@fnal.gov}}
\author{James Hirschauer}
\author{Ron Lipton}
\author{Benjamin Parpillon}\altaffiliation[Also at ]{University of Illinois Chicago, Chicago, IL 60607, USA}
\author{Chinar Syal}
\author{Nhan Tran}\altaffiliation[Also at ]{Northwestern University, Evanston, IL 60208, USA}
\affiliation{Fermi National Accelerator Laboratory, Batavia, IL 60510, USA}

\author{Petar Maksimovic}
\author{Morris Swartz}
\affiliation{Johns Hopkins University, Baltimore, MD 21218, USA}

\author{Nick Manganelli}
\affiliation{Northeastern University, Boston, MA 02115,
USA}

\author{Arghya Ranjan Das}
\author{Shiqi Kuang}
\author{Mia Liu}
\author{Ana Sof\'ia Calle Mu\~noz}
\altaffiliation[Also at ]{EAFIT University, Medel\'in, Antioquia, Colombia}
\affiliation{Purdue University, West Lafayette, IN 47907, USA}

\author{Daniel Abadjiev}
\author{Karri DiPetrillo}
\author{Eliza Howard}
\author{Rachel Kovach-Fuentes}  
\altaffiliation[Now at ]{Rice University, Houston, TX 77005, USA}
\author{Eric You}
\affiliation{University of Chicago, Chicago, IL 60637, USA}

\author{Jannicke Pearkes}
\author{Ricardo Silvestre}
\altaffiliation[Also at ]{University of Illinois Chicago, Chicago, IL 60607, USA}
\author{Keith Ulmer} 
\affiliation{University of Colorado Boulder, Boulder, CO 80309, USA}

\author{Harshul Gupta}
\author{Corrinne Mills} 
\author{Danush Shekar} 
\author{Amit Trivedi}
\author{Mohammad Abrar Wadud}
\affiliation{University of Illinois Chicago, Chicago, IL 60607, USA}

\author{David Jiang}
\author{Mark S. Neubauer}
\affiliation{University of Illinois Urbana-Champaign, Champaign, IL 61801, USA}

\date{\today}

\begin{abstract}
Modern scientific instruments operate under increasingly extreme constraints on bandwidth, latency, and power. Inference at the sensor edge determines experimental data collection efficiency by deciding which information to save for further analysis. Particle tracking detectors at the Large Hadron Collider exemplify this challenge: pixelated silicon sensors generate rich spatiotemporal ionization patterns, yet most of this information is discarded due to data-rate limitations. Concurrently, advancements in co-design tools provide rapid turnaround for incorporating machine learning into application-specific integrated circuits, motivating designs for particle detectors with new integrated technologies. We demonstrate that neural networks embedded in the front-end electronics can infer charged-particle kinematic parameters from a single silicon layer. We regress hit positions and incident angles with calibrated uncertainties, while satisfying stringent constraints on numerical precision, latency, and silicon area. Our results establish a path toward probabilistic inference directly at the edge, opening new opportunities for intelligent sensing in high-rate scientific instruments.
\end{abstract}

\maketitle


\section{Introduction}

Modern high-energy physics (HEP) experiments study the fundamental building blocks of matter by analyzing high-rate data that requires real-time inference under strict latency and bandwidth constraints. The multipurpose ATLAS \cite{Aad:1129811} and CMS \cite{Chatrchyan:1129810} experiments at CERN's Large Hadron Collider (LHC) generate petabytes of data per second, but the majority is discarded by real-time filtering systems, known as triggers, which consist of both specialized processing hardware and commodity computing. The unique latency, radiation, power, and data rate constraints define a machine learning (ML) regime fundamentally different from conventional edge or embedded inference. Our proposed methods reduce an array of on-sensor charge information to a fixed-length set of kinematic variables, thereby achieving rate reduction not possible with traditional methods.

Like most accelerator-based HEP experiments, ATLAS and CMS employ silicon tracking detectors, or trackers, in the region closest to the proton-proton (pp) collision point. Trackers provide precise measurements of each outgoing charged particle by sampling points along its trajectory (track), which can be used to reconstruct its point of origin (vertex). Track and vertex information constitute critical input to flagship measurements targeted by the LHC and high-luminosity LHC (HL-LHC)~\cite{atlastdr,cmstdr} programs. 

The high particle density in the trackers combined with the number of detector channels -- $\mathcal{O}$(100 million), to be upgraded to $\mathcal{O}$(2 billion) for the HL-LHC -- results in a data rate that far exceeds the bandwidth available for transfer and analysis at the pp collision frequency of 40\;MHz. The trigger must therefore rely on other detector subsystems to determine which collisions to save for further processing. While future trigger systems will incorporate limited tracking information \cite{Ryd_2020}, data from the pixel detector, the central and most finely segmented portion of the tracker, remain inaccessible to real-time inference due to bandwidth constraints. The increased granularity of next-generation pixel detectors will exacerbate this problem by further increasing the overall volume of data produced by the detector. 

In this work, we demonstrate the feasibility of on-detector reduction of pixel data using ML algorithms embedded directly in the front-end readout chip, or ASIC. A mixture density network~\cite{mdn_paper} is trained on the ionization signature of a charged particle in a single pixel layer to regress properties that are directly relevant for track and vertex reconstruction, such as particle hit position and incident angle. Variations on this model explore the impact of different architectures and number of outputs on both physics and hardware performance. All models must satisfy strict area, latency, and power constraints. Model performance is studied as such constraints are introduced, and candidate algorithms are synthesized for the 28~nm CMOS process using high level synthesis (HLS)~\cite{coussy2009high}. The performance of the regression models exceeds that of traditional reconstruction methods. 

The proposed regression models reduce an array of on-sensor charge information to a fixed-length set of kinematic variables, thereby decoupling the volume of output data and the geometry of the pixel sensor. This approach will enable future experiments to capitalize on the increased precision afforded by smaller pixels with little cost to the total bandwidth. This would unlock pixel detectors for use in high-rate real-time trigger and processing systems, which has not yet been achieved in HEP experiments.

\section{Related Work}
\label{sec:related_work}

On-detector data reduction is a central requirement for high-rate tracking detectors at the LHC.
The CMS Phase-2 Outer Tracker addresses this challenge by locally correlating hits in closely spaced sensor modules to form track stubs consistent with charged particles whose transverse momentum exceeds a programmable threshold, thereby reducing the data volume transmitted to the Level-1 trigger system~\cite{CMSPhase2TrackerTDR,Ryd_2020}.
In parallel, modern pixel readout ASICs, such as those developed by the RD53 collaboration, provide highly optimized front-end amplification, digitization, buffering, zero suppression, and high-speed readout under the radiation and bandwidth constraints of the HL-LHC environment~\cite{RD53}.
These systems establish the feasibility and necessity of local processing in detector front-end electronics, but they do not exploit the full charge pattern within a pixel cluster to infer local track parameters at the sensor edge.

Earlier work on intelligent tracking detectors explored moving data reduction and pattern recognition closer to the sensor through advanced integration technologies.
Vertically integrated pixel readout devices demonstrated that multiple tiers of electronics could be combined with fine-grained pixel arrays to form compact, highly integrated detector modules~\cite{Deptuch2010VIP}.
Related studies of three-dimensional integrated tracker modules proposed local correlations between closely spaced sensor layers to provide momentum discrimination before off-detector transmission~\cite{Lipton2010IntelligentTrackers}.
These developments established the broader intelligent-tracker concept: using integration and local logic to reduce data at the point where it is produced.
We build on this foundation by replacing deterministic local reductions with learned inference that maps the full pixel-cluster charge pattern to continuous track-parameter estimates and their uncertainties.

More recent smartpixels studies have applied ML to reduce pixel-detector data at the sensor edge.
There, neural network classifiers are trained on pixel-cluster charge patterns to reject hits from low-momentum particles before readout, and the associated readout architecture integrates charge-sensitive amplification, in-pixel digitization, and embedded digital neural network logic in 28 nm CMOS~\cite{Yoo2024SmartPixelSensors,Parpillon2024SmartPixels}.
Alternative approaches based on neuromorphic computing have also been explored for compact on-sensor transverse-momentum filtering~\cite{Kulkarni2023Neuromorphic}.
These works establish ML as a viable mechanism for on-detector filtering.
This work extends that direction from binary or threshold-based filtering to calibrated regression of local track parameters and their uncertainties.

The charge distribution of a pixel cluster has also been studied in the context of limited-precision readout.
Prior work showed that a small number of digitized charge bits can preserve much of the information needed for hit efficiency, position resolution, cluster classification, and particle identification in HL-LHC pixel detectors~\cite{Chen2018ChargeInformation}.
Charge sharing is likewise central to traditional local hit reconstruction in the CMS pixel detector, where projected cluster shapes and boundary-pixel charges are used to estimate the position of a hit~\cite{pixellocalreco}.
These studies motivate low-precision charge measurements for local reconstruction, but they rely on fixed reconstruction rules or offline information, whereas we learn a task-specific mapping from the full pixel-cluster charge pattern to local track parameters, constrained by the front-end ASIC.

Together, these studies show that local processing is essential for future high-rate tracking detectors, and that increasingly sophisticated logic can be integrated into pixel readout ASICs.
Existing approaches reduce data primarily by exploiting local correlations, applying fixed reconstruction rules, or making binary filtering decisions.
We instead use the pixel-cluster charge pattern to regress continuous charged-particle parameters directly at the sensor edge.
The probabilistic formulation additionally provides a calibrated uncertainty for each prediction, so that the reduced on-detector representation preserves information relevant for downstream track and vertex reconstruction.

\section{Simulated dataset}
\label{sec:dataset}

\begin{figure}[!htb]
    \centering
    \includegraphics[width=0.45\textwidth]{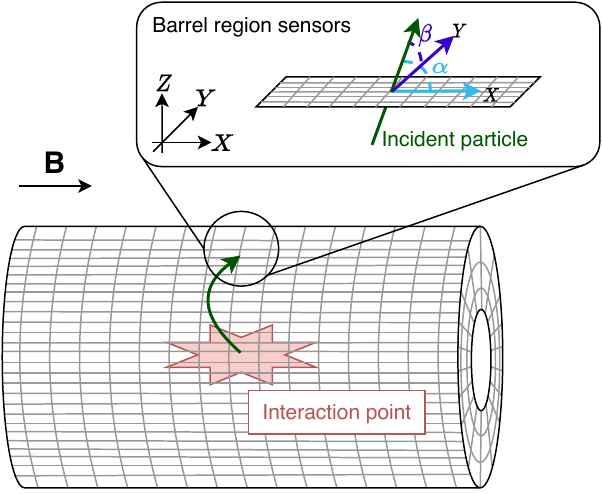}
    \caption{The silicon sensor is described by a Cartesian coordinate system in which the $x$-axis is parallel to the LHC beam direction, the $y$-axis lies in the sensor plane and is orthogonal to $x$, and the $z$-axis is normal to the sensor plane, pointing outward from the interaction region through the depleted sensor volume. The incident angles $\alpha$ and $\beta$ are defined with respect to this coordinate system.}
    \label{fig:coordinates}
\end{figure}

The spatial distribution and time evolution of charge deposited in a pixel array are sensitive to the traversing particle's kinematic properties. A particle's trajectory at the mid-plane of the pixel sensor is described by the following variables in the coordinate system of the sensor, shown in Figure \ref{fig:coordinates}: the hit position $x,y$ where the particle traverses the sensor; the polar incident angle $\alpha$; and the azimuthal incident angle 
$\beta$. 

The ML models described in this work are trained on simulated datasets of charged pion interactions in an array of pixels implemented in high resistivity silicon \cite{zenodo}. Electric field maps are generated using Silvaco TCAD \cite{silvaco_tcad} for a futuristic sensor geometry. The simulated sensor is composed of an array of $16\times16$ pixels, oriented in the $xy$ plane, with pitch 50 $\mu$m in $x$ and 12.5 $\mu$ in y. The thickness of the simulated sensor is 100 $\mu$m, and its volume is depleted of charge carriers by a reverse bias of -100 V. The sensor is immersed in a 3.8 T magnetic field oriented parallel to the $x$-axis.

Charged pions are generated at a distance of 30 mm from the silicon sensor. Positively and negatively charged pions are generated in equal numbers. The interaction of the charged pions with the sensor and subsequent charge collection and signal induction are simulated using PixelAV~\cite{pixelav} using the electric field maps described above. Charge is collected as a cumulative sum in each pixel over a time window of 4~ns in 200~ps increments. 

Only charge clusters that are fully contained in the simulated pixel array are considered: clusters with a total charge of 50 electrons or more deposited in the boundary pixels of the $16\times16$ pixel matrix are discarded. The analysis of clusters extending beyond the boundary of this array is left to future work. 

Two angular ranges are simulated. The first, used for training, places no constraints on how far in $x$ and $y$ a sensor module can extend. The resulting large ranges for both $\alpha$ and $\beta$ are limited only by the criterion that all clusters be contained in a $16\times16$ array. This first dataset of 520,000 clusters is partitioned into 80\% and 20\% for training and validation respectively. The second angular range assumes a sensor width (in $y$) of 16~mm, which restricts $\beta$ to a more realistic range. This dataset is used to evaluate the performance of the  neural network models. 



\section{Regression models \label{sec:models}}

To evaluate the feasibility of multiple implementation options, a suite of neural network algorithms is trained to predict charged particle $x$, $y$, $\cot\alpha$, and $\cot\beta$. For each model, three variants are trained to study how on-chip resource utilization depends on the details of the readout:
\begin{enumerate}
\item The \textbf{Max} model predicts values for $x$, $y$, $\cot\alpha$, $\cot\beta$, and the covariance matrix, for a total of 14 outputs. 
\item The \textbf{Full} models predict values for $x$, $y$, $\cot\alpha$, $\cot\beta$, and the uncertainty of one standard deviation on each variable ($\sigma_v$ for $v\in \{x, y, \cot\alpha, \cot\beta\}$), for a total of 8 outputs. 
\item The \textbf{Slim} models predict values for $x$, $y$, and $\cot\beta$ only, for a total of three outputs.
\end{enumerate}
The Max model provides the most information about the incident particle track, and the Full models provide nearly the same information since the off-diagonal elements of the covariance matrix are close to zero. 
The Slim models focus on a limited set of parameters selected for their potential to benefit a hardware-based track trigger. Because they output the least data, the Slim models are most likely to meet LHC bandwidth constraints. Because the Max and Full models allow for detailed error analysis, they provide critical validation of Slim model predictions.

All models are implemented in TensorFlow~\cite{tensorflow2015-whitepaper} using the Keras API \cite{chollet2015keras}.
Neural network training is performed on batches of 5000 clusters each for 1000 epochs with no early stopping. For each run, we retain the model weights corresponding to the best validation performance observed across the entire training history. All models are trained using the Adam optimizer \cite{adamoptimizer} with Nesterov Momentum incorporated \cite{dozat.2016} and a learning rate of $10^{-3}$. 

For the Max and Full models, we employ a mixture density network (MDN)~\cite{mdn_paper} to describe the possible parent distributions of reconstructed tracks. The response and resolution distributions of reconstructed tracks are inherently multi-Gaussian, but safely approximated as Gaussian per-cluster, from the varying number of hits per cluster.
Therefore, the network predicts the parameters of a single multi-dimensional Gaussian, and we construct a loss function from the mixture of the MDN at the level of the likelihood over the training data. The predicted means and standard deviations of the multidimensional Gaussian correspond to the regressed track parameters and uncertainties. We compare these uncertainties to residuals and confirm that our model is learning well-calibrated predictions. For the Slim models, which predict parameters without uncertainties, a mean-squared-error (MSE) loss function is used. 

Three model architectures are tested: one is based on 2D convolutional layers (Conv2D), the second is based on 1D convolutional layers (Conv1D), and the third is a multi-layer perceptron (MLP). 

\begin{figure*}
\centering

\begin{subfigure}[t]{0.25\textwidth}
    \begin{tikzpicture}[node distance=2cm]

\tikzset{
  act/.style={
    draw,
    fill=green!20,
    minimum height=0.55cm,
    inner sep=0pt,
    font=\scriptsize,
    align=center,
    text centered,
    text height=1.6ex,
    text depth=0.4ex
  }
}

\node (start) [startstop] {2D images};

\node (dw) [subconv, below=1cm of start] {DepthwiseConv2D};
\node (pw) [subconv, below=2mm of dw] {PointwiseConv2D};

\node (titlepad) [coordinate, above=4mm of dw.north] {};

\begin{scope}[on background layer]
  \node (layer1) [sepconv, fit=(titlepad)(dw)(pw), inner sep=4pt,
    label=right:{\scriptsize\shortstack[l]{\texttt{fixed<4,1>}}}
  ] {};
\end{scope}

\path (layer1.north) -- (dw.north) coordinate[pos=0.55] (seplabel);
\node[font=\small] at (seplabel) {SepConv2D};

\node (act1) [act, fit=(layer1.south west)(layer1.south east), yshift=-0.35cm] {};
\node[font=\scriptsize] at (act1.center) {\(\tanh\)};

\node (layer2) [conv, below=2mm of act1,
  label=right:{\scriptsize\shortstack[l]{\texttt{fixed<4,1>}}}
] {Conv2D};

\node (act2) [act, fit=(layer2.south west)(layer2.south east), yshift=-0.35cm] {};
\node[font=\scriptsize] at (act2.center) {\(\tanh\)};

\node (layer3) [pooling, below=2mm of act2] {AvgPool};

\node (layer4) [dense, below=2mm of layer3,
  label=right:{\scriptsize\shortstack[l]{\texttt{fixed<8,1>}}}
] {Dense};

\node (act3) [act, fit=(layer4.south west)(layer4.south east), yshift=-0.35cm] {};
\node[font=\scriptsize] at (act3.center) {\(\tanh\)};

\node (layer5) [dense, below=2mm of act3,
  label=right:{\scriptsize\shortstack[l]{\texttt{fixed<8,1>}}}
] {Dense};

\node (act4) [act, fit=(layer5.south west)(layer5.south east), yshift=-0.35cm] {};
\node[font=\scriptsize] at (act4.center) {\(\tanh\)};

\node (layer6) [dense, below=2mm of act4,
  label=right:{\scriptsize\shortstack[l]{\texttt{fixed<8,1>}}}
] {Dense};

\node (act5) [act, fit=(layer6.south west)(layer6.south east), yshift=-0.35cm] {};
\node[font=\scriptsize] at (act5.center) {\(\tanh\)};

\node (stop) [startstop, below=2mm of act5] {Outputs};

\draw [arrow] (start.south) -- (layer1.north);
\draw [arrow] (act1.south) -- (layer2.north);
\draw [arrow] (act2.south) -- (layer3.north);
\draw [arrow] (layer3.south) -- (layer4.north);
\draw [arrow] (act3.south) -- (layer5.north);
\draw [arrow] (act4.south) -- (layer6.north);
\draw [arrow] (act5.south) -- (stop.north);

\end{tikzpicture}
    \caption*{(a) Conv2D}
\end{subfigure}\hfill
\begin{subfigure}[t]{0.35\textwidth}
    \begin{tikzpicture}[node distance=2cm]

\tikzset{
  act/.style={
    draw,
    fill=green!20,
    minimum height=0.55cm,
    inner sep=0pt,
    font=\scriptsize,
    align=center,
    text centered,
    text height=1.6ex,
    text depth=0.4ex
  }
}

\node (start) [startstop] {2D images};

\node (layer1) [pooling, below of=start, yshift=0.75cm, xshift=-1.7cm] {Avg. Pool};
\node (start1) [startstop, below of=layer1, yshift=0.75cm] {$x$ projections};

\node (layer2) [conv, below of=start1, yshift=0.75cm,
  label=right:{\scriptsize\shortstack[l]{\texttt{fixed<4,1>}}}
] {Conv1D};

\node (layer3) [pooling, below of=start, yshift=0.75cm, xshift=1.5cm] {Avg. Pool};
\node (start2) [startstop, below of=layer3, yshift=0.75cm] {$y$ projections};

\node (layer4) [conv, below of=start2, yshift=0.75cm,
  label=right:{\scriptsize\shortstack[l]{\texttt{fixed<4,1>}}}
] {Conv1D};

\node (layer5) [concat, below of=layer4, yshift=0.75cm, xshift=-1.5cm,
  label=right:{\scriptsize\shortstack[l]{\texttt{fixed<4,1>}}}
] {Concatenate};

\node (act1) [act, fit=(layer5.south west)(layer5.south east), yshift=-0.35cm] {};
\node[font=\scriptsize] at (act1.center) {\(\tanh\)};

\node (layer6) [dense, below=2mm of act1,
  label=right:{\scriptsize\shortstack[l]{\texttt{fixed<8,1>}}}
] {Dense};

\node (act2) [act, fit=(layer6.south west)(layer6.south east), yshift=-0.35cm] {};
\node[font=\scriptsize] at (act2.center) {\(\tanh\)};

\node (layer7) [dense, below=2mm of act2,
  label=right:{\scriptsize\shortstack[l]{\texttt{fixed<8,1>}}}
] {Dense};

\node (act3) [act, fit=(layer7.south west)(layer7.south east), yshift=-0.35cm] {};
\node[font=\scriptsize] at (act3.center) {\(\tanh\)};

\node (layer8) [dense, below=2mm of act3,
  label=right:{\scriptsize\shortstack[l]{\texttt{fixed<8,1>}}}
] {Dense};

\node (act4) [act, fit=(layer8.south west)(layer8.south east), yshift=-0.35cm] {};
\node[font=\scriptsize] at (act4.center) {\(\tanh\)};

\node (stop) [startstop, below=2mm of act4] {Outputs};

\draw [arrow] (start) -- (layer1);
\draw [arrow] (start) -- (layer3);
\draw [arrow] (layer1) -- (start1);
\draw [arrow] (layer3) -- (start2);
\draw [arrow] (start1) -- (layer2);
\draw [arrow] (start2) -- (layer4);
\draw [arrow] (layer2) -- (layer5);
\draw [arrow] (layer4) -- (layer5);
\draw [arrow] (act1.south) -- (layer6.north);
\draw [arrow] (act2.south) -- (layer7.north);
\draw [arrow] (act3.south) -- (layer8.north);
\draw [arrow] (act4.south) -- (stop.north);

\end{tikzpicture}
    \caption*{(b) Conv1D}
\end{subfigure}\hfill
\begin{subfigure}[t]{0.35\textwidth}
%
%

\begin{tikzpicture}[node distance=2cm]

\tikzset{
  act/.style={
    draw,
    fill=green!20,
    minimum height=0.55cm,
    inner sep=0pt,
    font=\scriptsize,
    align=center,
    text centered,
    text height=1.6ex,
    text depth=0.4ex
  }
}

\node (start) [startstop] {2D images};

\node (layer1) [pooling, below of=start, yshift=0.75cm, xshift=-1.7cm] {Avg. Pool};
\node (start1) [startstop, below of=layer1, yshift=0.75cm] {$x$ projections};

\node (layer2) [dense, below of=start1, yshift=0.75cm,
  label=right:{\scriptsize\shortstack[l]{\texttt{fixed<8,1>}}}
] {Dense};

\node (actx) [act, fit=(layer2.south west)(layer2.south east), yshift=-0.35cm] {};
\node[font=\scriptsize] at (actx.center) {\(\mathrm{ReLU}\)};

\node (layer3) [pooling, below of=start, yshift=0.75cm, xshift=1.5cm] {Avg. Pool};
\node (start2) [startstop, below of=layer3, yshift=0.75cm] {$y$ projections};

\node (layer4) [dense, below of=start2, yshift=0.75cm,
  label=right:{\scriptsize\shortstack[l]{\texttt{fixed<8,1>}}}
] {Dense};

\node (acty) [act, fit=(layer4.south west)(layer4.south east), yshift=-0.35cm] {};
\node[font=\scriptsize] at (acty.center) {\(\mathrm{ReLU}\)};

\node (layer5) [concat, below=0.5cm of acty, xshift=-1.5cm,
  label=right:{\scriptsize\shortstack[l]{\texttt{fixed<8,1>}}}
] {Concatenate};

\node (layer6) [dense, below=2mm of layer5,
  label=right:{\scriptsize\shortstack[l]{\texttt{fixed<8,1>}}}
] {Dense};

\node (act2) [act, fit=(layer6.south west)(layer6.south east), yshift=-0.35cm] {};
\node[font=\scriptsize] at (act2.center) {\(\tanh\)};

\node (layer7) [dense, below=2mm of act2,
  label=right:{\scriptsize\shortstack[l]{\texttt{fixed<8,1>}}}
] {Dense};

\node (act3) [act, fit=(layer7.south west)(layer7.south east), yshift=-0.35cm] {};
\node[font=\scriptsize] at (act3.center) {\(\tanh\)};

\node (layer8) [dense, below=2mm of act3,
  label=right:{\scriptsize\shortstack[l]{\texttt{fixed<8,1>}}}
] {Dense};

\node (act4) [act, fit=(layer8.south west)(layer8.south east), yshift=-0.35cm] {};
\node[font=\scriptsize] at (act4.center) {\(\tanh\)};

\node (stop) [startstop, below=2mm of act4] {Outputs};

\draw [arrow] (start) -- (layer1);
\draw [arrow] (start) -- (layer3);
\draw [arrow] (layer1) -- (start1);
\draw [arrow] (layer3) -- (start2);

\draw [arrow] (start1) -- (layer2);
\draw [arrow] (actx.south) -- ([xshift=-0.2cm]layer5.north);

\draw [arrow] (start2) -- (layer4);
\draw [arrow] (acty.south) -- ([xshift=0.2cm]layer5.north);

\draw [arrow] (layer5.south) -- (layer6.north);
\draw [arrow] (act2.south) -- (layer7.north);
\draw [arrow] (act3.south) -- (layer8.north);
\draw [arrow] (act4.south) -- (stop.north);

\end{tikzpicture}
    \caption*{(c) MLP}
\end{subfigure}
\caption{Diagram of the architecture and bit precision for each layer of (a) the Conv2D, (b) the Conv1D, and (c) the MLP models.}
\label{fig:diagram}
\end{figure*}

\subsection{Conv2D architecture}
\label{sec:conv2d-models}

The Conv2D models take as input two-dimensional arrays of collected charge. As shown in Fig.~\ref{fig:diagram}a, the network begins with two convolutional stages that extract spatial features from the 2D charge image, then applies average pooling to reduce the spatial resolution and parameter count, and finally uses three dense layers to generate the regression outputs.

The convolutional front-end of the model begins with a two-dimensional depthwise-separable convolution (\texttt{SeparableConv2D}), implemented as a depthwise spatial convolution followed by a $1\times1$ pointwise convolution. This decomposition substantially reduces the number of parameters and multiply-accumulate operations compared to a standard \texttt{Conv2D} layer, while maintaining sensitivity to local spatial structure in the charge footprint~\cite{howard2017mobilenets,chollet2017xception}. In our implementation, the depthwise stage uses a $3\times3$ kernel and the pointwise stage projects the result into 5 output feature maps. The separable convolution is followed by a hyperbolic tangent activation ($\tanh$). A second convolutional stage then applies a standard \texttt{Conv2D} layer with 5 filters and a $1\times1$ kernel, followed by another $\tanh$ activation. Finally, an average pooling layer (\texttt{AvgPool}) reduces the spatial resolution of the feature maps, limiting the size of the subsequent dense layers and reducing the complexity of a hardware implementation.

After the pooling stage, the feature maps are flattened and passed to a three-layer fully connected stage. The first two dense layers contain 16 nodes each and are followed by $\tanh$ activations. The third and final dense layer produces the regression output, with 14, 8 or 3 nodes for the Max, Full, and Slim Conv2D configurations, respectively.

Finally, to mitigate overfitting and stabilize the learned response to fluctuations in the collected charge patterns, we apply \texttt{L1L2(0.01)} regularization to the layer weights and constrain intermediate activations using an \texttt{L2(0.01)} activity penalty.

\subsection{Conv1D architecture}
\label{sec:conv1d-models}
The Conv1D models take as input two-dimensional arrays of collected charge and build one-dimensional summaries by constructing the $x$- and $y$-projections through a pair of initial average pooling branches. Because these projections collapse one dimension, they discard spatial information in the charge footprint. As a result, this architecture is not suitable for the Max configuration, which additionally regresses the full covariance matrix and therefore requires modeling correlations between $x$, $y$, $\cot\alpha$, and $\cot\beta$ in a consistent way. Each projection is then processed by its own 1D convolutional layer, and the resulting feature sequences are concatenated and passed to a three-layer fully connected block to produce the regression outputs. A diagram of the Conv1D architecture is shown in Fig.~\ref{fig:diagram}b.

Concretely, the \(x\)-projection is obtained via \texttt{AveragePooling2D} with pooling window \((1,16)\), averaging along the \(y\) direction while preserving the \(x\) coordinate. The \(y\)-projection is obtained with pooling window \((16,1)\), averaging along the \(x\) direction while preserving \(y\). After pooling, each projection is reshaped into a sequence of length 16, retaining the time-slice dimension as channels. Each branch is then processed by a \texttt{Conv1D} layer with \(5\) filters and kernel size \(3\). The two branches are concatenated along the sequence dimension and passed through a \(\tanh\) activation. The output is flattened and passed to two dense layers with \(16\) nodes and \(\tanh\) activations, followed by a final dense layer with \(8\) or \(3\) outputs for the Full and Slim configurations, respectively.

\subsection{MLP architecture}
\label{sec:mlp-models}

The MLP models take as input two-dimensional arrays of collected charge and replace the convolutional front-end with a lightweight projection stage followed by the same fully connected regression block described in Sec.~\ref{sec:conv1d-models}. Specifically, the input is summarized into an \(x\)-projection and a \(y\)-projection using two \texttt{AveragePooling2D} branches; each projection is flattened and passed through a dense layer that maps it into a compact latent representation. We use a \texttt{ReLU} activation in these projection branches to avoid early saturation effects that can occur with bounded nonlinearities (e.g., \(\tanh\)). The two latent vectors are concatenated and processed by an additional dense layer with \(\texttt{hidden}=16\) units and a \(\tanh\) nonlinearity, which forms the final embedding.

The embedding is then passed to a fully connected three-layer regression stage consisting of two dense hidden layers with 16 nodes each and \(\tanh\) activations, followed by a linear output layer. The output dimension is set to 8 for the Full configuration and to 3 for the Slim configuration.  As in the previous architectures, we apply \texttt{the L1L2(0.01)} regularization to the layer weights and an \texttt{L2(0.01)} activity penalty to intermediate activations to improve generalization and stabilize the learned response to fluctuations in the charge patterns.

\section{Input charge digitization with end-to-end training \label{sec:digi}}

The analog front-end in the proposed ASIC uses a charge-sensitive amplifier (CSA)~\cite{csa_gatti} followed by a flash analog-to-digital converter (ADC) to convert the charge collected in each pixel to a two-bit discrete representation. This discretization, referred to as digitization, is emulated in model training by partitioning the full range of charge into four bins. The collected charge per pixel in the simulated dataset is digitized by assigning each pixel the corresponding bin index: 
\begin{equation}
    \label{eq:SoftDig-piecewise}
    D(\mathcal{Q}_\text{in})=
    \begin{cases}
    0 & \text{if } -\infty < \mathcal{Q}_\text{in} < T_0 \\
    1 & \text{if } T_0 \leq \mathcal{Q}_\text{in} < T_1 \\
    2 & \text{if } T_1 \leq \mathcal{Q}_\text{in} < T_2 \\
    3 & \text{if } T_2 \leq \mathcal{Q}_\text{in} < +\infty
    \end{cases}
\end{equation}
where thresholds $\mathcal{T} = \{-\infty,T_0,T_1,T_2,\infty\}$ correspond to the boundaries of the selected charge bins.  The threshold values can be configured on-chip by adjusting bias voltages. 

The choice of thresholds determines what information about the charge cluster is passed to the regression network, which is implemented in digital logic downstream of the ADC. We perform end-to-end training in which the optimal thresholds are learned jointly with the inference task. A dedicated neural network is used to extract the optimal thresholds by relating them to trainable parameters in a custom \textit{SoftQuantize} layer. This layer is prepended to the nominal regression models and performs the function of digitization by the ADC in training. The sole purpose of this layer is to learn the optimal thresholds ($\mathcal{T}$), and it is not intended for implementation on-ASIC. This approach allows the network to discover a digitization scheme that preserves task-relevant information. Conceptually, this shifts digitization from a fixed engineering choice to a learned component of the inference pipeline.

\subsection{Formulation of SoftQuantize}

Physically meaningful thresholds must be strictly ordered, i.e. $T_{j-1} < T_j$ for all $j$. Instead of defining $\mathcal{T}$ directly, we therefore define a set of unconstrained trainable parameters $\theta$, which are transformed into strictly positive intervals $\Delta$:
\begin{equation}
    \Delta_i = \ln(1 + e^{\theta_i}),
\end{equation}
The physical thresholds are then constructed as the cumulative sum of these positive intervals, starting from a lower bound $T_{\text{min}}$, as
    \begin{equation}
        T_j := T_{\text{min}} + \sum_{k=0}^{j} \Delta_k.
    \end{equation}
This transformation guarantees monotonicity in $\mathcal{T}$ and $T_j > T_{\text{min}}$ for any real-valued $\theta$.

Standard ``hard'' quantization functions, such as Equation \ref{eq:SoftDig-piecewise}, employ non-differentiable step functions that block any gradient flow required during backpropagation. To optimize the thresholds and model weights via gradient-descent, we instead introduce a differentiable SoftQuantize layer to perform the function of the ADC during training. 

The SoftQuantize layer employs a Straight-Through Estimator (STE)-like strategy using hard quantization $D$ (Equation \ref{eq:SoftDig-piecewise}) during the forward pass and a differentiable formulation, $D_\text{soft}$, for gradient approximation during the backward pass. We define the ``soft''
quantization  using the logistic sigmoid function $\sigma$: 
\begin{equation}
    \label{eq:SoftDig-soft_dig}
    D_\text{soft}(x) := \sum_{j=0}^{2^B - 1 } \sigma\left(\frac{k\cdot (T_j - x)}{\tau_j}\right).
\end{equation}

Two key parameters control the behavior of the soft quantization: the local scale factor $\tau_j$ and dimensionless global temperature $k$. The local scale factor is used to make the sigmoid functions insensitive to the relative bin width. For each threshold, $\tau_j$ is defined as
    \begin{equation}
        \tau_j := \frac{\Delta_j + \Delta_{j+1}}{2}.
    \end{equation}
    
Importantly, scaling by $\tau_j$ also makes the temperature $k$ dimensionless and \textit{global} across the sigmoids, allowing a single variable to control the overall steepness. Larger values of $k$ produce steeper sigmoids, and in the limit of infinite temperature, the sigmoids converge to step-functions. To leverage this behavior during optimization, the temperature $k$ is annealed over the course of training. The temperature is initiated with a small value $k_\text{init}\simeq1$ and slowly increased to a larger value $k_\text{max}\simeq67$ as the training progresses. A cosine annealing schedule is used to control the annealing. 

\subsection{Selection of thresholds}

The choice of thresholds in a pixel detector is driven by the amount of intrinsic sensor and electronic noise, with the lowest threshold typically corresponding to an upward fluctuation of five standard deviations $(\sigma)$. In the input data for the threshold optimization training, an additional charge $\epsilon$ is injected to reflect detector noise. The total charge collected in each pixel becomes
    \begin{equation}
        \label{eq:SoftDig-Noise_Sig}
         \mathcal{Q}_\text{in} = \mathcal{Q}_\text{sim} + \epsilon, 
    \end{equation}
where $\mathcal{Q}_\text{sim}$ is the simulated charge resulting from a traversing particle and $\epsilon$ is randomly sampled from a Gaussian distribution $\mathcal{N}(0, \sigma_{\text{noise}})$. The value of $\sigma_{\text{noise}}$ is estimated to be 80 electrons based on simulations designed in the Cadence Virtuoso flow and simulated with Spectre \cite{cadence}. 

\begin{figure*}[!htb]
    \centering
    \includegraphics[width=1\textwidth]{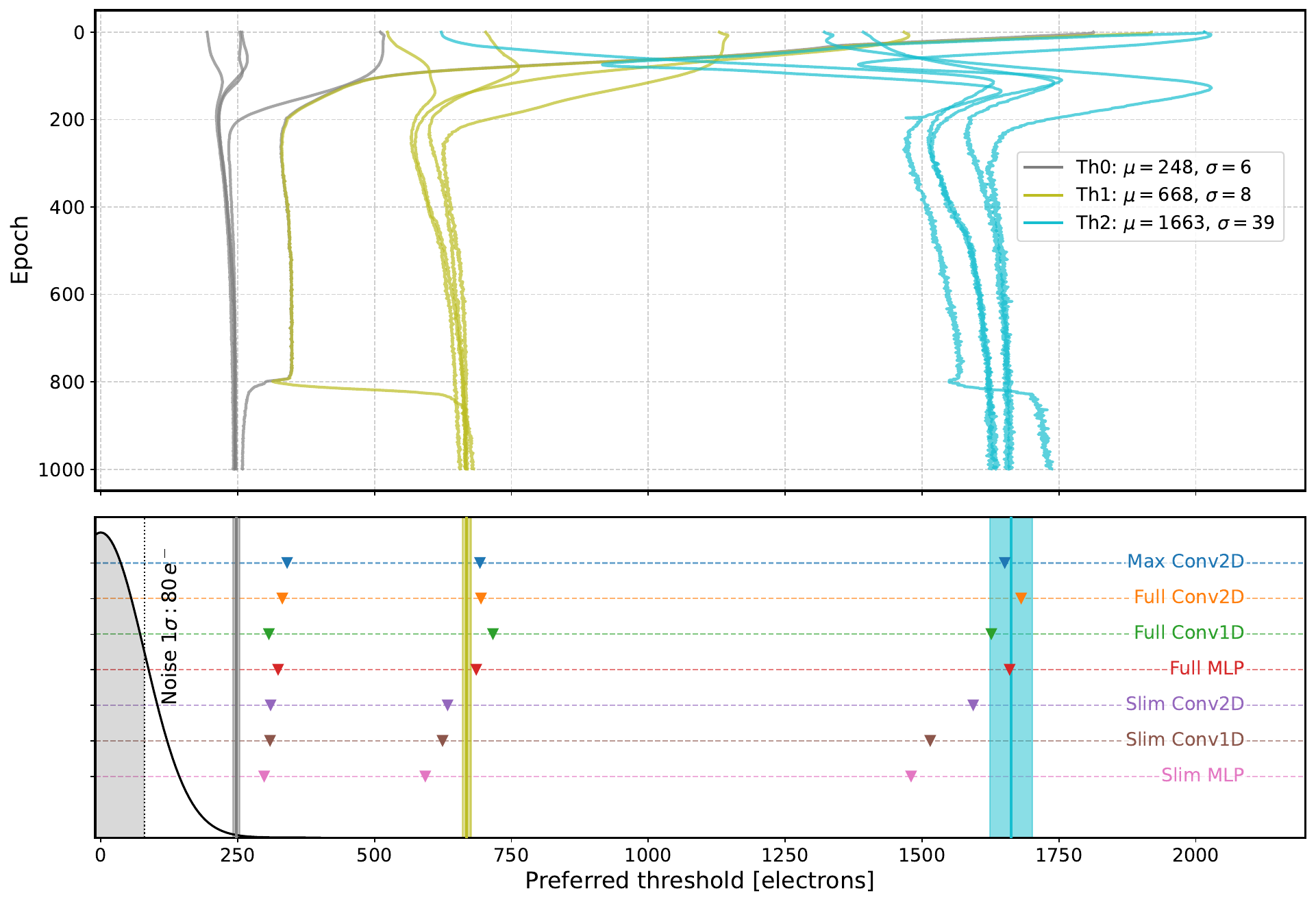}
    \caption{Top: Threshold values as a function of epoch in the training of the Max transformer with SoftQuantize.  Bottom: Points indicate the preferred charge thresholds for each regression model. The vertical lines and shaded bands represent, respectively, the means and standard deviations of the thresholds obtained by the transformer. }
    \label{fig:Threshold_Summary}
\end{figure*}

The optimal thresholds are determined by training the Max model with a SoftQuantize layer on the noisy input data. A high-capacity transformer architecture~\cite{vaswani2023attentionneed,dosovitskiy2021imageworth16x16words} is chosen for this task to ensure robust convergence of the charge thresholds. The transformer was trained multiple times with random initial thresholds, as shown in the upper panel of Figure \ref{fig:Threshold_Summary}. The model converges to approximately the same set of thresholds regardless of initial values. The mean and standard deviation of the learned thresholds are displayed in Figure \ref{fig:Threshold_Summary}.

The mean thresholds determined by the transformer were subsequently used as the initial values for trainings of the resource-constrained models with the SoftQuantize layer on the noisy input data. The preferred thresholds for each model,  shown in the lower panel of Figure \ref{fig:Threshold_Summary}, are overall close to those determined by the transformer. The optimal minimum threshold for all models is higher than that selected by the transformer, but below the $5\sigma$ level at which most pixel detectors operate. 

\section{Results \label{sec:results}}

We demonstrate for the first time that ML methods can regress 
particle hit positions and incident angles from ionization patterns in a single tracking detector layer. We also present comprehensive studies of model variants that explore the trade-off between information content, uncertainty quantification, and hardware feasibility when inference is pushed to the sensor edge.

Each model presented is trained in four configurations corresponding to different levels of realism in hardware implementation and data preparation. These configurations, ordered from most idealized to most realistic, are:
\begin{enumerate}
\item Training data consisting of simulated charge collected at \textit{20 time frames} separated by 200 ps, each with electron-level precision; network weights and activations have 32-bit floating-point precision. These models serve as an indicator of optimal performance. 
\item Training data consisting of simulated charge collected at \textit{two time frames} separated by 3.8 ns, each with electron-level precision; network weights and activations have 32-bit floating-point precision.
\item Training data consisting of simulated charge collected at two time frames separated by 3.8 ns, each with \textit{two-bit precision} and using the optimal thresholds shown in the lower panel of Figure \ref{fig:Threshold_Summary}; network weights and activations have 32-bit floating-point precision.
\item Training data consisting of simulated charge collected at two time frames separated by 3.8~ns, each with two-bit precision; \textit{convolutional weights and biases are quantized} using \texttt{fixed<4,1>}, while dense-layer weights and biases use \texttt{fixed<8,1>}. Activation and intermediate precisions are selected separately through profiling. This is the model variant synthesized in Section~\ref{sec:asic}, where the precision choices are described in detail.
\end{enumerate}

\begin{figure*}[!htb]
    \centering
    \includegraphics[width=1\textwidth]{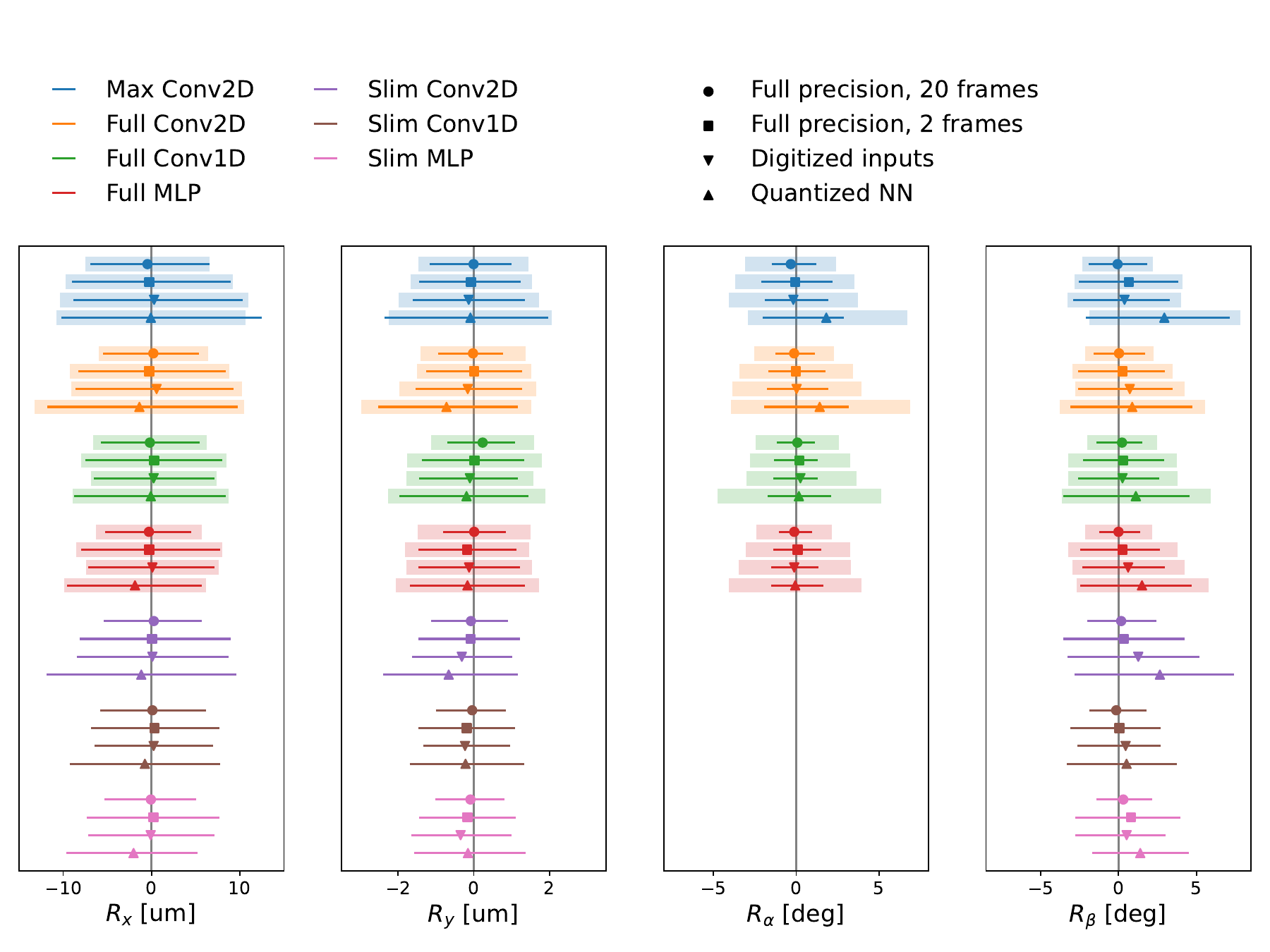}
    \caption{Comparison of the residuals for  $x$, $y$, $\alpha$, and $\beta$ for all models. The solid lines represent the minimum interval containing 68\% of clusters. For the Max and Full models, the shaded bands represent the mean predicted uncertainty on each variable. Across architectures, the dominant performance degradation arises from temporal sparsification of the input, whereas aggressive charge quantization introduces comparatively minor loss.}
    \label{fig:residuals}
\end{figure*}

A summary of the residuals $R_v = v - v_\text{true}$ for $v\in \{x, y, \alpha, \beta\}$ is shown in Figure \ref{fig:residuals} for all models. The points represent the mean of the residual distribution ($\bar{R}_v$), and the solid lines represent the minimum interval containing 68\% of clusters in the test set, $I_v(68\%)$. The Conv2D models have the widest residual distributions, and therefore the poorest resolution, for most parameters, while the Conv1D and MLP models achieve similar performance. The residual distributions have little dependence on the number of outputs. 

For the Max and Full models, where an uncertainty $\sigma_v$ is predicted for each variable $v$, the mean regressed uncertainty is represented by the shaded bands in Figure \ref{fig:residuals}. For $x$, $y$, and $\beta$, the predicted $\sigma_v$ is close to but greater than $I_v(68\%)$, indicating that the model slightly over-estimates the uncertainty on average. The mean predicted uncertainty on $\alpha$, which all models significantly over-estimate, is dominated by clusters from charged particles with $\alpha$ close to $90^\circ$. These charge clusters span only a few pixels in the $x$-direction, providing very little shape information for the neural network to extract and resulting in large $\sigma_\alpha$. The value of $\sigma_\beta$ is more accurately estimated because clusters close to $\beta=90^\circ$ span more pixels due to the smaller pixel pitch in the $y$-direction. 

The progression from idealized to realistic models reveals which hardware constraints have the largest impact on physics performance. Although models operating on many time frames of input data cannot be implemented on-ASIC due to power constraints, the time evolution of the charge cluster is clearly rich in physics content: most models trained with two time frames have 30-40\% worse resolution than their counterparts trained on 20 time frames, with a few of the angular predictions degrading by more than 60\%. 
Applying the optimal thresholds for two-bit digitization typically results in only 5-10\% worse resolution than training the same models with electron-level precision. Finally, the quantization of the neural network parameters introduces a shift in the mean of the residual distribution for some output features, with the Conv2D models experiencing the largest bias.

\subsection{Parameter estimation without ML \label{sec:non-ml}}

For the sake of comparison, $x$, $y$, $\alpha$, and $\beta$ are also calculated using non-ML methods. The integrated cluster charge after 4 ns, which corresponds to the final time frame in the simulated dataset, is digitized with the optimal thresholds obtained by the transformer (Section \ref{sec:digi}) and used as input to these algorithms. The performance of the non-ML algorithms is compared to the MLP Full model with two-bit input digitization and quantized weights and biases (red upward triangles in Figure \ref{fig:residuals}). Comparison to the MLP Slim model yields similar conclusions for $x$, $y$, and $\beta$.

\begin{figure}[!htb]
    \centering
    \includegraphics[width=0.45\textwidth]{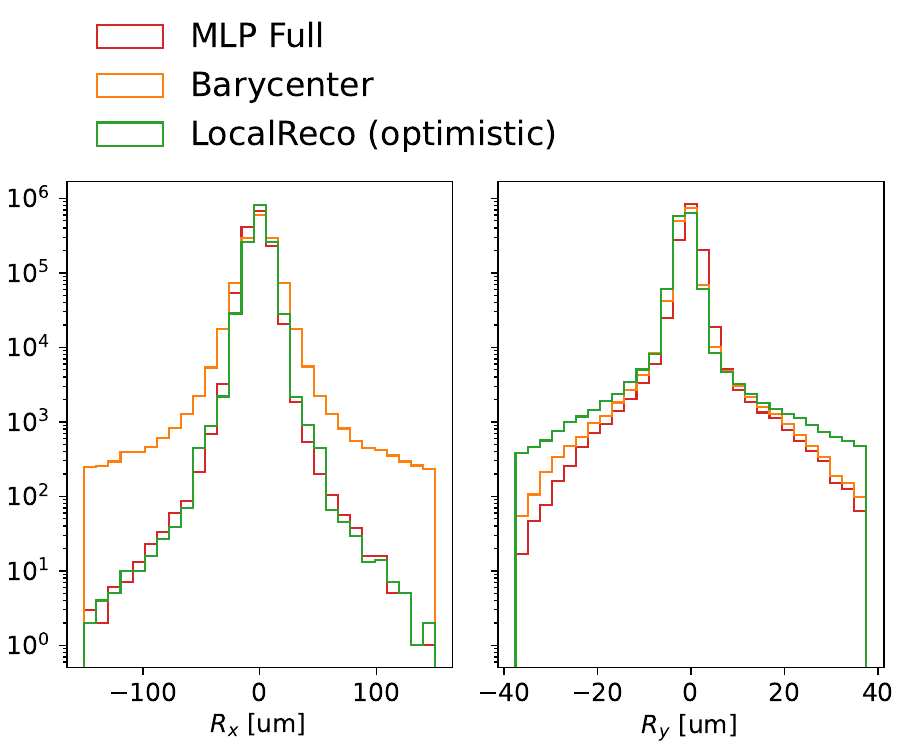}
    \caption{Comparison of the residuals for $x$ and $y$ between the Full MLP model and non-ML algorithms. The non-ML algorithms labeled as ``optimistic'' incorporate information that is not directly available on the ASIC.}
    \label{fig:noMLxy}
\end{figure}

The $x$ and $y$ predictions of the MLP Full model are compared to two non-ML algorithms: Barycenter and LocalReco. The Barycenter algorithm uses a weighted average of the charge cluster projected onto the respective axis. The value $x_\text{bary}$ obtained from this weighted average corresponds directly to the predicted $x$ position, while an additional correction is applied to $y_\text{bary}$ to account for the deflection of charge carriers by the magnetic field (Lorentz drift). The predicted $y$ position is
\begin{align}
    y = y_\text{bary} + \frac{\Delta_y}{2}
\end{align}
where $\Delta_y = T\tan\Theta_L$ for sensor thickness $T$ and Lorentz angle $\Theta_L$.

The LocalReco algorithm corresponds to the local reconstruction that CMS employs offline to determine the position of a hit in the pixel detector~\cite{pixellocalreco}. Like the Barycenter algorithm, LocalReco calculates $x$ and $y$ from the charge cluster projected onto the respective axis. The LocalReco prediction for the $x$ position of a hit is
\begin{align}
\label{eq:xlocalreco}
x = \frac{Q_L^x - Q_F^x}{Q_L^x + Q_F^x}\cdot\frac{|T\cot\alpha| - (x_L-x_F)}{2} + \frac{x_F + x_L}{2},
\end{align}
where $Q_F^x$ ($Q_L^x$) is the amount of charge collected in the first (last) pixel of the charge projection, and $x_F$ ($x_L$) is the $x$-coordinate of the boundary between the first and second (next-to-last and last) pixels in the projection. The LocalReco prediction also depends on the value of $\alpha$, which in offline analysis would be determined from a track fit to multiple pixel layers. For the comparison presented here, perfect knowledge of $\alpha$ is assumed.  The LocalReco prediction for the $y$ position of a hit is analogous to Equation \ref{eq:xlocalreco} with an additional correction for Lorentz drift:
\begin{eqnarray}
y = && \frac{Q_L^y - Q_F^y}{Q_L^y + Q_F^y}\cdot\frac{|T\cot\beta + \Delta_y| - (y_L-y_F)}{2} \nonumber \\ && + \frac{y_F + y_L}{2} + \frac{\Delta_y}{2} .
\end{eqnarray}
Here, $Q_F^y$ ($Q_L^y$) is the amount of charge collected in the first (last) pixel of the charge projection, and $y_F$ ($y_L$) is the $y$-coordinate of the boundary between the first and second (next-to-last and last) pixels in the projection.  As above, $\Delta_y$ is a correction accounting for Lorentz drift.

Figure \ref{fig:noMLxy} shows the $x$ and $y$ residuals for the MLP Full model and the Barycenter and LocalReco algorithms. Comparing the performance of the Full MLP model to LocalReco yields a striking result: the simple on-ASIC network can estimate $x$ and $y$ from a single pixel layer with comparable accuracy to the offline reconstruction, which relies on multiple pixel layers. The ML model also significantly outperforms the Barycenter calculation for both $x$ and $y$. Both non-ML methods introduce a large average bias in the $y$ parameter compared to the ML model: $\bar{R}_y = -0.70$ and $-0.81$ microns for Barycenter and LocalReco, respectively, compared to only $-0.1$ microns for MLP Full.

\begin{figure}[!htb]
    \centering
    \includegraphics[width=0.45\textwidth]{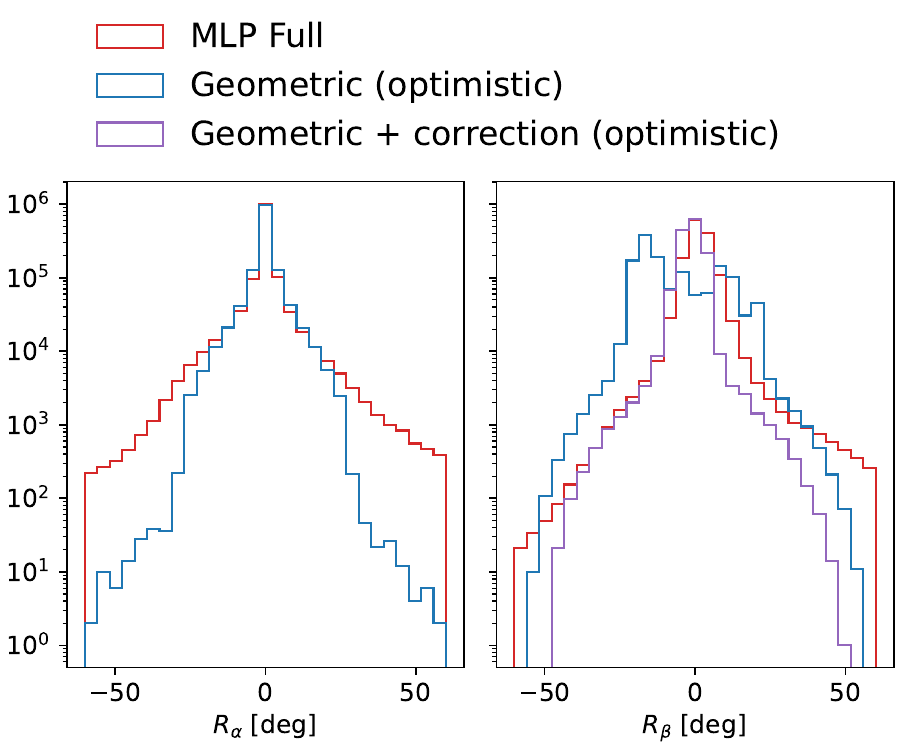}
    \caption{Comparison of the residuals for $\alpha$ and $\beta$ between the Full MLP model and non-ML algorithms. The non-ML algorithms labeled as ``optimistic'' incorporate information that is not directly available on the ASIC.}
    \label{fig:noMLab}
\end{figure}

The $\alpha$ and $\beta$ residuals for the MLP Full model are compared to a non-ML Geometric method. The Geometric method for estimating $\alpha$ uses the size of the charge cluster projected onto the $x$-axis:
\begin{align}
|\cot\alpha | = (w_x-1) \; \frac{p_x }{T},
\end{align}
where $w_x$ is the size of the projected charge (in pixels), $p_x$ is the pixel $x$-pitch, and $T$ is the sensor thickness. 
This provides an estimate of the absolute value of the angle, and perfect knowledge of the sign of $\alpha$ is assumed for this comparison. For charge clusters spanning multiple pixels in the $x$-direction, the sign of $\alpha$ can be reliably determined by comparing the two time frames used as input to the training, but this is an optimistic assumption for small clusters. An analogous Geometric calculation is performed to estimate $\beta$ from the $y$-projected cluster size, but is found to perform poorly without any correction for Lorentz drift. A more precise calculation of $\beta$ applies a correction to account for changes in the size of the cluster $y$-projection due to Lorentz drift:
\begin{eqnarray}
|\cot\beta| = && (w_y-1) \; \frac{p_y}{T} \nonumber \\ && + \begin{cases}
    T\tan\Theta_L, & \beta < \pi + \Theta_L\\
    0, & \beta = \pi + \Theta_L\\
    -T\tan\Theta_L, & \beta > \pi + \Theta_L
\end{cases}
\end{eqnarray}
The sign of the Lorentz drift correction depends on how $\beta$ compares to $\pi + \Theta_L$, which cannot be well estimated geometrically from the charge cluster. For the comparison presented here, perfect knowledge of both the sign of $\beta$ and the sign of $\beta - \pi - \Theta_L$ (highly idealized) is assumed. 

Figure \ref{fig:noMLab} shows the $\alpha$ and $\beta$ residuals for the MLP Full model and the Geometric algorithms. Again the comparison between the MLP Full model and the non-ML algorithm is striking: the ML model obtains better performance in both $\alpha$ and $\beta$ than the Geometric method, despite the inclusion of highly optimistic assumptions in the Geometric calculation. The ML model gives better $\alpha$ resolution in the core of the distribution, but exhibits longer tails than the Geometric method. The Geometric $\beta$ calculation without the Lorentz drift correction (blue) exhibits poor performance.  The Full MLP model gives similar resolution to the Lorentz-drift corrected geometric algorithm (purple) and has a lower average bias: $\bar{R}_\beta = +0.6^\circ$ compared to $-1.1^\circ$.

\section{ASIC synthesis \label{sec:asic}}

We followed a standard codesign workflow to translate the neural networks described in Section \ref{sec:models} into low-latency hardware~\cite{Fahim:2021hls4ml}.  The floating-point TensorFlow/Keras model implementations described were used as the initial model definition and training environment, providing a reference implementation for the subsequent quantized design. Then we ran quantization-aware training using QKeras to introduce fixed-point numerical precision as a trainable design choice while minimizing accuracy loss~\cite{Coelho_2021}. Finally, we used \texttt{hls4ml} to translate the trained quantized model into C++ for high-level synthesis, enabling an iterative optimization loop between physics performance and hardware metrics such as latency, throughput, and resource usage~\cite{vloncar_2021_5680908,Duarte:2018ite}.

Figure~\ref{fig:diagram} reports the fixed-point format used for weights and biases on the right-hand side of each layer. We use the notation \texttt{fixed<W,I>}, where $W$ is the total bit width and $I$ is the number of integer bits, including the sign bit. The weights and biases of the convolutional layers were quantized using \texttt{fixed<4,1>}, corresponding to a representable range of $[-1, 1-1/8]$ with resolution $1/8$, while the weights and biases of the dense layers used \texttt{fixed<8,1>}, corresponding to a representable range of $[-1, 1-1/128]$ with resolution $1/128$. These formats were selected empirically to preserve physics performance with only limited degradation relative to the floating-point baselines.

The trained QKeras models are translated into synthesizable C++ using \texttt{hls4ml}. The generated code is synthesized with Siemens Catapult HLS~\cite{catapult-hls}, targeting a TSMC 28nm technology node with a clock period of 25~ns. Designs are fully unrolled to achieve an initiation interval of one clock cycle, enabling continuous inference at each LHC bunch crossing. Profile-guided optimization is used to determine the bit widths of the accumulator and intermediate buffer, imposing a mean squared error per-layer below $10^{-4}$ relative to the QKeras fixed-point reference. After HLS, we perform validation and confirm that the hardware implementation preserves the regression performance of the quantized model.

Table~\ref{tab:synthesis_results} summarizes the HLS synthesis results for all models. All designs achieve a latency of two clock cycles, corresponding to $50~\mathrm{ns}$, and an initiation interval of one clock cycle, allowing one new inference every $25~\mathrm{ns}$ and meeting the system requirements.
The designs also meet the 25~ns timing constraint with substantial implementation headroom: the timing slack ranges from 14.33~ns to 16.62~ns, corresponding to a margin of 57--66\% of the clock period. This margin provides useful headroom for subsequent physical-design effects, including routing and clock-tree insertion. However, timing closure under worst-case operating conditions must be confirmed through post-layout analysis.
The Slim Conv2D implementation is the most compact overall at $0.2880~\mathrm{mm}^2$. The Full MLP implementation is substantially smaller than the Full convolutional models despite its higher parameter count, suggesting that implementation area depends strongly on the architecture and resulting hardware dataflow.

\begin{table*}[!htb]
    \small
    \centering
    \begin{tabular}{l | c | ccc | ccc}
    & Max models & \multicolumn{3}{c|}{Full models} & \multicolumn{3}{c}{Slim models}  \\
    & Conv2D & Conv2D & Conv1D & MLP & Conv2D & Conv1D & MLP \\ \hline
    NN Parameters & 1898 & 1767 & 2734 & 2264 & 1682 & 2649 & 2179 \\
    Latency / II [clk] & 2/1 & 2/1 & 2/1 & 2/1 & 2/1 & 2/1 & 2/1\\
    Slack [ns] & 14.33 & 14.33 & 14.33 & 16.36& 16.62 & 15.72 & 15.72\\
    HLS area [mm$^2$] & 0.6353 & 0.6197 & 0.6067 & 0.3004 & 0.2880 & 0.3487 & 0.3350 \\
    \hline
    \end{tabular}
    \caption{Model size, latency/initiation interval (II), slack, and area results for each model, derived from HLS targeting TSMC 28~nm and 25~ns clock period.}
    \label{tab:synthesis_results}
\end{table*}


\section{Conclusion \label{sec:conclusion}}

We have presented a suite of compact neural network algorithms that extract the kinematic features of a charged particle traversing one layer of a silicon tracking detector. Using the ionization signature in an array of pixels, these networks extract the incident particle's position and crossing angles, as well as uncertainties on these quantities. The networks are optimized for implementation on-ASIC and are synthesizable in 28~nm CMOS, meeting system requirements for area and latency.  This demonstrates a new capability of intelligent sensing that arises from embedding probabilistic ML directly into the data acquisition chain under extreme resource constraints. Moreover, we demonstrate the first joint optimization of signal digitization thresholds and neural network weights to achieve optimal physics performance. 

The performance of each model is evaluated based on the residual distributions of the predicted variables. The best performance of each model is obtained by training on the collected charge sampled at twenty 200 ps intervals, indicating that the time evolution of the charge cluster is rich in physics information. However, ASIC power requirements limit the sampling rate, and the models proposed for on-chip implementation operate on the collected charge at two time points. The two-bit digitization of the input charge and the quantization of the neural network weights and biases also result in worse resolution for all parameters, but the effect is small compared to the reduction of input data from twenty to two time frames. 

A selected model is compared to non-ML methods for reconstructing particle kinematics. The non-ML algorithms incorporate highly optimistic assumptions, including perfect knowledge of data external to the sensor. The ML model nevertheless outperforms all non-ML models while remaining sufficiently small and efficient for implementation within the pixelated area of an ASIC. The on-chip ML algorithms thus yield novel sensing capabilities.

This work illustrates how ML can be integrated at the earliest stages of data creation. It provides a concrete example of probabilistic inference and uncertainty estimation under extreme constraints on numerical precision, latency, and power, a regime that remains underexplored in mainstream ML benchmarks. 

The results highlight two complementary advances. First, physics quantities can be extracted directly at the sensor edge with an accuracy that matches or surpasses conventional reconstruction. Second, co-design tools and workflows for training, quantization-aware optimization, and hardware synthesis are readily adaptable to a wide range of scientific instruments. Similar constraints arise in cryogenic, space-based, and high-radiation environments, where end-to-end co-design of sensing, digitization, and inference offers a pathway toward intelligent instruments that maximize scientific return under fixed resource budgets.
\subsection{Limitations and future work} 
Several limitations of this study suggest directions for future work. The training dataset can be made more hardware-realistic by, for example, modeling the effects of the CSA and ADC on the collected charge cluster. We can explore structured pruning or sparsification to further reduce resource usage of the ML models. And although HLS demonstrates feasibility, full hardware validation remains an essential next step.
Our results indicate that the time evolution of charge carries significant information, and extending our models to incorporate richer time structure will be important as readout technologies evolve. 
\textcolor{black}{The results presented here are based on HLS synthesis, for which power estimates are approximate: pre-layout tools lack switching-activity annotation and omit clock distribution, local buffering, and the analog front-end and ADC. The selected architectures and numerical formats were chosen to limit expected digital power consumption; however, power compliance cannot be established from the present HLS study and will require post-layout characterization.
This study also omits radiation-aware synthesis, irradiation testing, and an evaluation of single-event upsets in stored configuration parameters and network weights, which are left as future work.
Nevertheless, the target $28\,\mathrm{nm}$ CMOS process has demonstrated favorable total ionizing dose tolerance in HL-LHC-oriented studies, and prototypes have been fabricated in the same node~\cite{parpillon2025inpixelintegrationsignalprocessing}.}
\textcolor{black}{
The present study also does not account for radiation damage to the sensor bulk, which alters the electric field profile and introduces charge trapping, such that the collected charge patterns evolve away from the unirradiated simulation used for training.
Since both the network weights and the digitization thresholds are reconfigurable on-chip, future work could explore periodic retraining on simulations of the irradiated sensor, with updated parameters deployed over the detector lifetime.
}
Finally, the co-design workflow developed in this work provides a valuable template for studying ML under extreme constraints beyond the specific application considered here.

\begin{acknowledgments}
The work presented in this document was conducted using the resources of the Fermi National Accelerator Laboratory (Fermilab), a U.S. Department of Energy (DOE), Office of Science, Office of High Energy Physics HEP User Facility. Fermilab is managed by Fermi Forward Discovery Group, LLC, acting under Contract No. 89243024CSC000002.

This work was completed using computing resources at the University of Chicago's Research Computing Center and the Fermilab Elastic Analysis Facility. We thank Burt Holzman (Fermilab) for computing support. 

We acknowledge the Fast Machine Learning collective as an open community of multi-domain experts and collaborators. This community, Javier Duarte and Vladimir Loncar in particular, were important for the development of this project.

We would like to extend our sincere gratitude to Harish Jamakhandi and David Burnette from Siemens EDA for their assistance and expertise with Catapult HLS.
 
ML and ARD are supported by NSF AREAS PHYS-2425206 and by the A3D3 HDR Institute through the NSF award PHY-2117997. ARD is also partially supported by the Fermilab Guest \& Visitor program.
MSN and DJ are supported through the NSF cooperative agreement OAC-2117997 and the DOE Office of Science award DE-SC0023365. 
RK was supported by the Metcalf Fellowship program of the University of Chicago.
ASCM was supported by the Undergraduate Research Experience Purdue - Colombia program.
CM, DS, HG, and MAW are supported by NSF award PHY-2208803 and, together with AT, by DOE Office of Science award DE-SC0023715 from Funding Opportunity Announcement for Artificial Intelligence Research for High Energy Physics, DE-FOA-0002705. DS is also supported by the University Research Alliance Visiting Scholar program.
JD and BW are supported by the DOE Early Career Research Program award DE-SC0026236.

DB, GDG, FF, AG, LG, JH, RL, BP, and CS are supported by Fermi Forward Discovery Group, LLC under Contract No. 89243024CSC000002 with the U.S. Department of Energy, Office of Science, Office of High Energy Physics. 
GDG and LG are partially supported under DOE project ``HAAI: Designing Smart Detectors with a ML-to-Silicon Platform'' (LAB 24-3305). 
GDG and BP are also supported by the U.S. Department of Energy, Office of Science, Offices of High Energy Physics and Advanced Scientific Computing Research, as part of the Co-design and Heterogeneous Integration in Microelectronics for Extreme Environments (CHIME) Microelectronics Science Research Center (MSRC) (DOE Lab Announcement No. LAB 24-3320).
%
%
FF, GDG, BP, and NT are also supported by the DOE Early Career Research Program. 

KFD and EH are supported by the NSF CAREER Program through award 2443370, and KFD is additionally supported by the Neubauer Family Assistant Professor Program. EY is supported by the University of Chicago's Quad Undergraduate Research Scholar program, and DA is supported by the University of Chicago's Sachs Fellowship. 

MS and PM are supported by NSF award PHY-2310072. 
NM is supported by the Fermilab LHC Physics Center (LPC) Distinguished Researcher award. 
JP and KU are supported by DOE award DE-SC0010005.
\end{acknowledgments}

\appendix*
\section{System design considerations \label{sec:system}}

In order to realize improvements from the proposed on-device pixel cluster reconstruction, it is necessary to consider the system that would be required to extract and analyze this data. Smartpixels generates a stream of reconstructed pixel clusters with positional and angular data at the rate of data creation.
The clearest use case for this data is to provide this information to a real-time (Level-One, L1) trigger required to control output data rates in an experiment.
Real-life examples of this can be found in the upgraded ATLAS or CMS detectors at the CERN HL-LHC, where CMS is building a L1 track trigger.
Introducing pixel information into these systems has the potential to greatly improve their performance, but that information must fit within the capabilities of experimental infrastructure such as cooling, powering, and data transfer.

\begin{table*}[!hbt]{
    \centering
    \begin{tabular}{|c|c|}
         \hline
         \textbf{Quantity} & \textbf{Value}\\
         \hline
         Phase 2  Max. Avg. Pixel-Cluster Density, 200 PU, $\mathrm{r}=3.3 ~\mathrm{cm}$ & 10 cm$^{-2}$  \\
         \hline
         Phase 2 Pixel Sensor Area & 3.61 cm$^2$ \\
         \hline
         HL-LHC Bunch Spacing & 25 ns \\
         \hline
        \begin{tabular}{@{}c@{}}Max. Avg. 40 MHz Bandwidth per Phase 2 sensor (+5 S.D.),\\ reading out active pixels \textit{no compression}\end{tabular} & 72.2 (132.2) Gbps \\
        \hline
        Smartpixels data reduction factor from \(p_T\) filter & 0.25 \\
         \hline
        Avg. number of active pixels per cluster, 200 PU, $\mathrm{r}=3.3~\mathrm{cm}$ & 5 \\
        \hline
        \begin{tabular}{@{}c@{}}Max. Avg. 40 MHz Bandwidth per smartpixels sensor (+5 S.D.),\\ 16 bits read out per cluster \textit{no compression}\end{tabular} & 5.77 (10.58) Gbps \\
        \hline
    \end{tabular}
    \caption{Inputs and calculations for the HL-LHC CMS pixel detector output data rate at 40 MHz.}\label{tab:smartpix_datarate}}
\end{table*}

Because a single inference engine processes one cluster per clock cycle, multiple engines or spatially distributed processing blocks would be required to sustain the full cluster rate of a sensor. The corresponding system-level area, routing, buffering, and power costs are not included in the single-model HLS results and will be addressed in future implementation studies.

To estimate how much cluster information we can send to a real-time trigger system we begin with the expected cluster flux upon the innermost layer of the CMS pixel detector: 10 clusters per $\mathrm{cm}^2$ per proton-proton bunch crossing and consider one bunch crossing.
This cluster density is derived from GEANT simulation with full out-of-time pileup simulated over a period of 20 bunch crossings.
Taking the average cluster yield per bunch-crossing we assume a 16-bit word is used to encode the physics information of each cluster to yield an average expected data rate.
While this 16-bit word does restrict the amount of information that can be saved per cluster, it is enough to encode meaningful position and bending angle information for reconstructed track stubs.
Then to estimate the maximum sustainable data rate in a streaming readout system we consider the five-standard-deviations upward fluctuation in the number of clusters compared to the expected Poisson mean. 
These calculations are summarized in Table~\ref{tab:smartpix_datarate}.
It is important to note that using the filtering and regression capabilities of smartpixels we are able to reduce the bandwidth required to read out a pixel sensor at the HL-LHC by a factor of 10.

Modern silicon photonics encoders operate at 30 Gbps and above~\cite{fastopticallinks}, and radiation hardness has been demonstrated for 30 Gbps encoders to HL-LHC expected doses.
Moreover, wavelength multiplexing can be employed to encode multiple streams on a single fiber, making 100 Gbps readout rates per sensor possible, or similarly yielding ultra-light readouts by using a single 30 Gbps encoder per sensor and stacking them on the same readout fiber.
This then implies that it is possible to stream this pixel data to real-time processing facilities like the CMS or ATLAS Level-One trigger systems and to then exploit the additional precise measurements made by the pixel detector at trigger level.
Our initial estimates of bandwidth therefore demonstrate that novel triggering strategies are made possible by smartpixels together with silicon photonics and the corresponding physics impacts should be studied.

\bibliography{bibliography}

\end{document}